\documentclass[%
 reprint, amsmath,amssymb,aps,prb]{revtex4-2}

\usepackage{amsmath}
\usepackage{physics}
\usepackage{graphicx}% Include figure files
\usepackage{dcolumn}% Align table columns on decimal point
\usepackage{bm}% bold math
\usepackage{hyperref}% add hypertext capabilities

\begin{document}

\preprint{APS/123-QED}

\title{Identifying the phases of Kane-Mele Hubbard Hamiltonian in momentum space: A many-body configuration interaction study}
%\thanks{A footnote to the article title}

\author{Ranadeep Roy$^{\dagger}$, Sasmita Mohakud$^{\ddagger}$, Katsunori Wakabayashi$^{\parallel}$ and Sudipta Dutta$^{*}$}
\affiliation{$^{\dagger}$Department of Physics, The Ohio State University, 191 West Woodruff Ave., Columbus, OH 43210, USA \\
$^{\ddagger}$Department of Physics, School of Advanced Sciences, Vellore Institute of Technology, Vellore, Tamil Nadu, 632014, India \\
$^{\parallel}$Department of Nanotechnology for Sustainable Energy, School of Science and Technology, Kwansei Gakuin University, Sanda, Hyogo 669-1330, Japan \\
$^{*}$Department of Physics, Indian Institute of Science Education and Research (IISER) Tirupati, Tirupati - 517619, Andhra Pradesh, India}
\email{sdutta@iisertirupati.ac.in}

\date{\today}% It is always \today, today,

\begin{abstract}
We investigate the magnetic and conduction properties of Kane-Mele Hubbard model in quasi one-dimensional honeycomb ribbon systems at half-filling by varying the strength of both spin-orbit interaction and on-site Coulomb correlation term. We use the numerical many-body configuration interaction (CI) method to investigate the dispersions of charge and spin gaps along with the momentum resolved spin-density profile over the full Brillouin zone. While the spin sector retains its topological nature at all values of spin-orbit coupling and Hubbard term, we report a new signature of the topological phase transition in the charge sector. This phase transition from a topological band insulating phase to a antiferromagnetically ordered Mott insulating phase is characterized by a shift of the many-body charge gap minima from Brillouin zone boundary to Dirac point. Our results provide a better understanding of the shifting of the gap-closing point in the momentum space which was reported in an earlier mean-field study of the same model and suggests an alternative numerical route to detect topological phase transition in strongly-correlated systems in terms of their momentum space behaviors.
\end{abstract}

\maketitle

%\tableofcontents

\section{\label{sec:level1}Introduction }

Topological insulators have garnered a lot of attention in the past two decades owing to their novel property of having an insulating bulk gap while hosting metallic edge states much like systems that exhibit integer quantum Hall effect and Chern insulators\cite{kane-mele1,haldane,moore,murakami,bernevig,konig}. However, unlike the later ones, topological insulators do not require the presence of magnetic field. The first microscopic realization of such materials, which are termed as $\mathbb{Z}_2$ topological insulators, was given by the Kane-Mele model in which the spin-orbit coupling mediated hopping in graphene and its nanoribbons gives rise to topological insulating behavior\cite{kane-mele1,kane-mele2,roy}.

Among the materials in which both spin-orbit coupling and electronic correlations play a dominant role, $\text{Na}_2\text{IrO}_3$\citep{shitade,laubach} and transition metal trichalcogenides ($\text{TMBX}_3$ : TM denotes transition metal,  B can be phosphorus/silicon/germanium and X denotes chalcogen) have generated significant interest in the recent years with several experiments reporting mono to few layer forms of the material\citep{sykim,chenggong,jaeunglee,chengtaikuo} and have also been the subject of several recent theoretical investigations\citep{archana,mishra}. This has motivated us to examine the effects of competition between onsite Coulomb correlation and spin-orbit coupling using Kane-Mele Hubbard model. In particular, we study the effects on the magnetic and conduction properties of quasi-one dimensional honeycomb systems via semi-empirical many-body configuration-interaction (CI) method. It provides us two main advantages. Firstly, there is no sign-problem arising from exchange of fermionic operators as it takes into account the relevant sign changes while generating the many-body basis set itself. Therefore, this method can even be used for other models which may not be sign-problem free and hence cannot be investigated via quantum Monte-Carlo techniques. Secondly, there is no restriction arising from the dimensionality of the system and hence it can be used to obtain reliable results for quasi-one dimensional systems\cite{sdutta-physrev,sdutta-scirep}.

Within the many-body CI framework, we investigate the momentum-resolved spin and charge degrees of freedoms of quasi-one-dimensional zigzag edge honeycomb nanoribbons (zHNRs) in presence of competing on-site Coulomb correlation and the spin-orbit interaction. Our study provides a new way of characterizing the quatum phases based on dispersion relations of spin and charge gaps and provides more generalized understandings of the existing mean-field and many-body results in momentum space. 

  \begin{figure}		
	\includegraphics[scale=0.30]{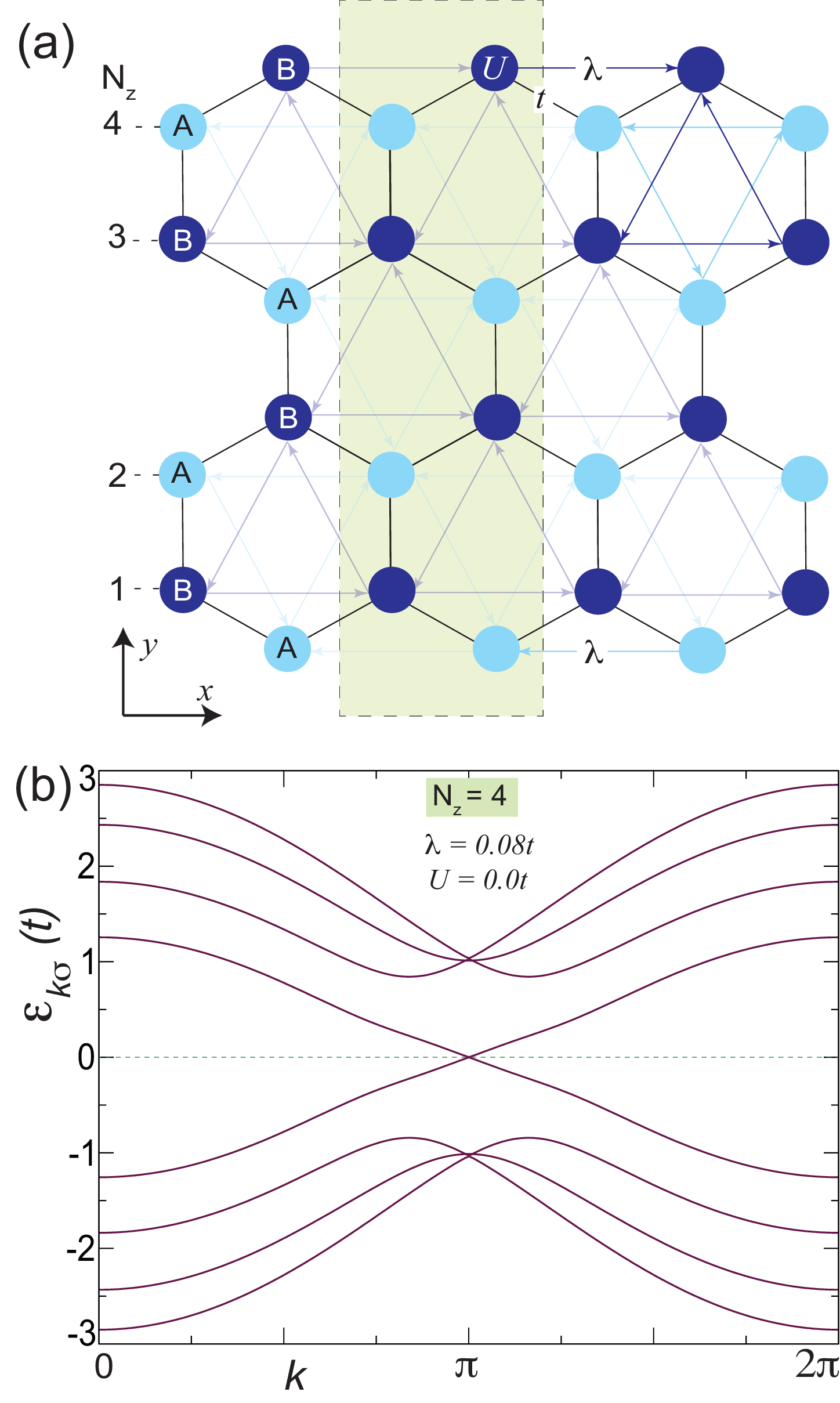}
	\caption{(Color onine) (a) The schematic representation of zigzag edge honeycomb nanoribbon of width $N_z=4$, modelled within Kane-Mele Hubbard Hamiltonian. The unit cell is depicted by the shaded rectangle, consisting of two different types of sublattice points A and B, marked with dark and light color circles. The spin-orbit coupling ($\lambda$) is introduced in terms of next nearest neighbor hoppings, marked by the arrows. (b) The energy dispersion of the Kane-Mele Hamiltonian for the same system, with $\lambda=0.08t$.}
  \end{figure}

The article is organized as follows. In section \ref{sec:level2}, we describe the model Hamiltonian and the system under consideration. We also give a brief overview of the many-body CI method adopted here. Section \ref{sec:level3} contains the main results of the work regarding the electronic and magnetic properties of our system where we also present a brief discussion of what is known about the bulk and the edge phase diagram of Kane-Mele Hubbard model. We conclude by summarising our main findings in section \ref{sec:level4}.

\section{\label{sec:level2}Model and Method \protect }

The Kane-Mele Hubbard Hamiltonian is defined as follows \cite{archana}:

\begin{align}\label{eq:1}
    & H \, = \, H_{KM} \, + \, H_{U} \nonumber \\
    & H_{KM} \, = \, -t\sum_{\langle \beta, \beta'\rangle} \sum_{\langle i,j\rangle \sigma}  c^{\dagger}_{\beta\sigma}(i)c_{\beta'\sigma}(j) \, \nonumber \\
	& + i\, \lambda \sum_{\langle \beta, \beta'\rangle}\sum_{\langle\langle i,j\rangle \rangle \sigma \tau}\nu_{ij}  c^{\dagger}_{\beta \sigma}(i) \, s^z_{\sigma \tau} \,c_{\beta' \tau}(j) \,  \nonumber \\ 
    & H_{U} \, = \, U\sum_{j} \, n_{j\uparrow}n_{j\downarrow}
\end{align}

\noindent where $c^{\dagger}_{\beta \sigma}(i) (c_{\beta \sigma}(i))$ creates (annihilates) an electron at site $i$ of unitcell $\beta$ with spin $\sigma$, $\langle \beta, \beta' \rangle$ and $\langle i,j\rangle$ are the nearest neighbor unitcells and nearest neighbor atomic sites in an unitcell, respectively, $\langle\langle i,j\rangle \rangle$ is the next nearest neighbor atomic sites, $s^z_{\sigma \tau}$ is the $z$-component of the Pauli matrix, $n$ is the number operator, $t$ is the hopping integral, $\lambda$ is the spin-orbit coupling strength and $U$ is the strength of onsite Coulomb interaction. $\nu_{ij} \, = \, \text{sgn}\left[ (\Vec{d}_i  \cross \Vec{d}_j)_z \right] \, = \, \pm 1$ determines the sign of the spin-orbit coupling mediated next nearest neighbour hopping term where $\Vec{d}_i$ and $\Vec{d}_j$ are the two vectors connecting sites $i$ and $j$. Note that, we do not consider any Rashba type spin-orbit coupling.

Our system consists of zHNRs with periodic boundary conditions along the $x$-direction as shown in Fig.1(a). The width of the system is described by the number of zigzag chains $(N_z)$ in a unit cell with alternate A and B sublattice points. We then apply the following Fourier transformations to the above Hamiltonian:

\begin{align}
 & c^{\dagger}_{\beta \sigma}(j) = \frac{1}{\sqrt{L_{x}}}\sum_{k}e^{-i\textbf{k}\textbf{r}_{\beta}}a^{\dagger}_{k\sigma}(j) \nonumber \\
 & c_{\beta \sigma}(j) = \frac{1}{\sqrt{L_{x}}}\sum_{k}e^{+i\textbf{k}\textbf{r}_{\beta}}a_{k\sigma}(j)
\end{align}

Finally we get the following momentum space Hamiltonian:

\begin{align}
    & H_{KM} = \sum_{k\sigma} \varepsilon_{k\sigma} a^{\dagger}_{k\sigma}(i) a_{k\sigma}(j) \nonumber \\
    & H_{U} = \frac{U}{L_{x}}\sum_{k,k',q}\sum_{j} a^{\dagger}_{k+q\uparrow}(j)a_{k\uparrow}(j)a^{\dagger}_{k'-q\downarrow}(j)a_{k'\downarrow}(j)
\end{align}

\noindent where $\varepsilon_{k\sigma}$ is the one-electron energy dispersion of $\sigma$ spin and $L_{x}$ is the length of the unit cell along $x$-direction.

The topological edge states can be obtained by solving the tight-binding part of the Hamiltonian for this strip geometry \cite{waka1,waka2,waka3,zarea}. Fig.1(b) shows the spin resolved band structure of $H_{KM}$ for a zigzag nanoribbon of width $N_z=4$ with degenerate up and down spin bands. Restricting our attention to any one spin sector at a time, one can see that there are two states traversing the Fermi level, which intersect at exactly $k=\pi$ and are localized at the opposite edges of the nanoribbon. Near $k=\pi$, the slope $(\partial_kE(k))$ is constant and the magnitude of the slope decreases with decrease in the strength of spin-orbit coupling $\lambda$ and when $\lambda$ becomes zero, one gets the partial flat-bands associated with zHNRs\cite{waka1,waka-ssp}. Further, the region in Brillouin zone, over which the edge bands exhibit linear behavior increases with increase in the strength of spin-orbit coupling. It should be noted that in zigzag nanoribbons the band-gap vanishes only for ribbons with odd number of zigzag chains in a unit cell \cite{chung}. For even number of zigzag chains, there is a small gap opening at $k=\pi$ which increases with the strength of spin-orbit coupling constant $\lambda$. However, for the small values of $\lambda$ taken into consideration in this work ($\lambda \leq 0.10t$), the gap can be practically taken to be zero.

To simplify the two-electron part of the Hamiltonian, i.e., $H_{U}$, we do not take into account any inter-$k$ scattering. Therefore, we consider $k = k'$ and $q = 0$. To solve the momentum space Kane-Mele Hubbard Hamiltonian, we adopt many-body CI method \cite{sdutta-physrev,sdutta-scirep}. The multi-determinant nature of many-body CI enables one to capture the effects of electronic interactions while the molecular orbital space approach allows one to make a natural set of approximations when dealing with larger systems. Therefore, we consider the following transformation to the molecular orbital space ($\mu$) by taking linear combination of the atomic orbitals:

\begin{align}
     & a^{\dagger}_{k\sigma}(j) = \sum_{\mu}\alpha^{*}_{j,\mu}(k)b^{\dagger}_{k\sigma}(\mu) \nonumber \\
	 & a_{k\sigma}(j) = \sum_{\mu}\alpha_{j,\mu}(k)b_{k\sigma}(\mu)
\end{align}

The coefficients $\alpha$ are obtained by solving the one-electron part of the Hamiltonian, $H_{KM}$ and are used to transform the two-electron part of the Hamiltonian, $H_{U}$ as follows,

\begin{widetext}
\begin{align}
    & H_{U} = \frac{U}{L_{x}}\sum_{k}\sum_{j}\sum_{\mu_{1},\mu_{2},\mu_{3},\mu_{4}} [\alpha^{*}_{j,\mu_{1},\uparrow}(k)\alpha_{j,\mu_{2},\uparrow}(k)\alpha^{*}_{j,\mu_{3},\downarrow}(k)\alpha_{j,\mu_{4},\downarrow}(k)] b^{\dagger}_{k\uparrow}(\mu_{1})b_{k\uparrow}(\mu_{2})b^{\dagger}_{k\downarrow}(\mu_{3})b_{k\downarrow}(\mu_{4})
\end{align}
\end{widetext}

In this study, we conduct a thorough study of the properties of zHNRs with $N_z \, = \, 4$ and $20$ and choose to present the results of $N_z \, = \, 4$ in the following sections. In case of $N_z \, = \, 4$, one can take all the molecular orbitals and hence carry out full-CI calculations. One should note that the total electron number operator $(N_{\text{tot}})$ commutes with the Kane-Mele Hubbard Hamiltonian. Also, absence of Rashba term means that the $z$-component of the total spin, i.e., $S_z^{\text{tot}}$ continues to be a good quantum number and thus allows us to reduce the size of the Hilbert space by restricting to specific $N_{\text{tot}}$ and $S_z^{\text{tot}}$ sector. For example, with $N_z \, = \, 4$ at half-filling one has 4900 basis states when one fixes $S_z^{\text{tot}} \, = \, 0$. The above approximation of same-$k$ scattering and consideration of all possible excitations within the chosen energy window around the Fermi energy to construct the many-body CI Hamiltonian matrix are known to show good agreement with the mean-field results and experimental observations on graphene nanoribbons\cite{sdutta-scirep,makarova-scirep,baringhaus-jpcm}. Such numerical approach allows us to explore the momentum-resolved properties of zHNRs.

To check the consistency of our results, we also analyze wider nanoribbons of width $N_z \, = \, 20$ in which we truncate the many-body configuration space by considering complete active space (CAS) with four valence bands and four conduction bands around the Fermi energy. By restricting to a small energy window near Fermi level, one can efficiently capture the low energy properties of the system in consideration.

\section{\label{sec:level3}Results \protect}

Before we present our results, we briefly review what is known about the bulk and edge-phase diagram of KMH model from several numerical and mean-field studies \citep{Hohenadler11,Hohenadler12,Rachel,Soriano,Lado,meng}. When the spin-orbit coupling is switched off, the system exhibits three phases : semi-metallic (SM) phase at small values of $U/t$ ($0$ to $3\sim 4$), quantum spin liquid (QSL) phase at intermediate values of $U/t$ ($4$ to $5$) and antiferromagnetically ordered Mott insulating (AFMI) phase at large value of $U/t$ ($> 5$). When spin-orbit coupling $\lambda$ is turned on, the gapless SM phase vanishes and one has topological band insulating (TBI) state at low value of $U/t$. The QSL phase is found to be stable for small values of $\lambda/t$ and at large values of $U/t$ there is a transition to $xy$-AFMI state. It has been found for the nanoribbon systems that the spin gap remains zero at all values of $U/t$ with $\lambda=0.1t$\cite{Yamaji}. On the other hand, the charge gap analysis shows a transition from gapless to gapped excitations at $U/t=3$. In case of $2D$ system, both the charge and the spin gaps remain non-zero upto $U/t=7$. Beyond this critical value, the spins exhibit gapless excitations while the charge gap continues to remain open. This led to the identification of three distinct phases at $\lambda=0.1t$: TBI phase for $U/t$ between $0$ to $3$, topological edge Mott insulator for $U/t$ between $3$ to $7$ and bulk antiferromagnetic Mott insulator (BAFI) for $U/t$ greater than $7$. Although the existing literature indicates the overall phase diagram of Kane-Mele Hubbard model for honeycomb systems, the microscopic understanding of the momentum-resolved spin and charge gap behavior arising from the interacting spins under the influence of competing spin-orbit coupling and electronic correlation within many-body formalism is still lacking. In this article, we present detailed analysis of such aspects.

  \begin{figure}		
	\includegraphics[scale=0.32]{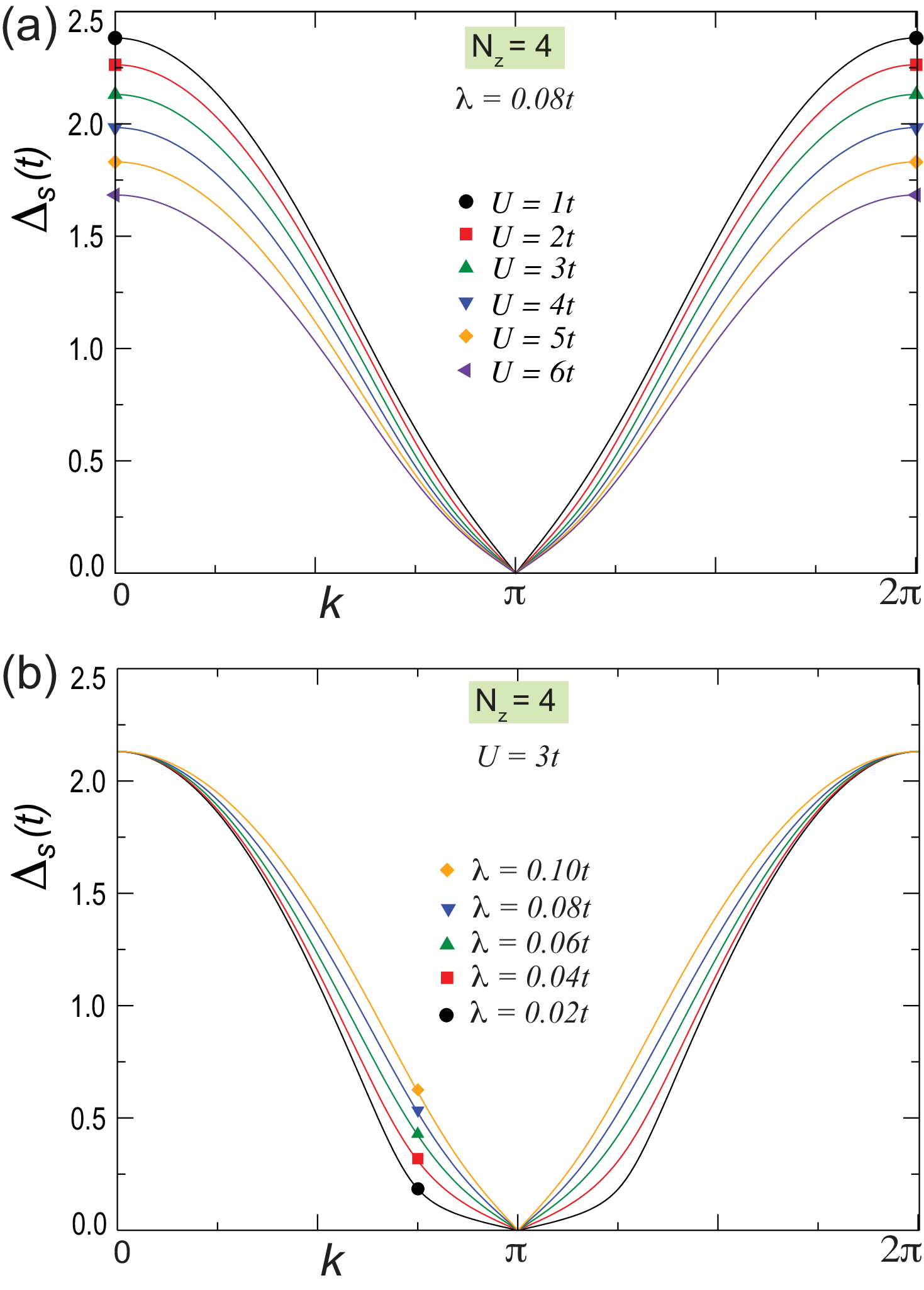}
	\caption{(Color online) The variation of spin gap ($(\Delta_s(k))$) as a function of momentum for a zigzag edge honeycomb nanoribbon with $N_z=4$ for (a) a fixed value of $\lambda=0.08t$ with $U/t$ varying from $1$ to $6$ and (b) a fixed value of $U/t=3$ with $\lambda$ varying from $0.02t$ to $0.1t$.}
  \end{figure}

\subsection{Spin gap}

To understand the magnetic properties of the system, we compute the spin-excitation gap $(\Delta_s)$ which is the energy required to flip one spin. At half-filling, this is given by the expression :

\begin{equation}
    \Delta_s(k) \, = \, \left[ E_0^N(S_z^{\text{tot}}=1) \, - \, E_0^N(S_z^{\text{tot}}=0)\right](k)
\end{equation}

\noindent where $E_0^N(S_z^{\text{tot}})$ denotes the ground state energy of the system with $N$ number of electrons having $z$-component of the total spin $S_z^{\text{tot}}$. Note that for zHNRs, one has $N \, = \, 2 \, N_z$ at half-filling. Fig.2(a) shows the momentum-resolved spin gap behavior for a fixed value of $\lambda=0.08t$ with varying $U$ and Fig.2(b) shows the same for a fixed value of $U=3$ with varying $\lambda$. We find the usual suppression in spin gap throughout the Brillouin zone with increase in the strength of on-site Coulomb interaction term (see Fig.2(a)). This can be attributed to higher spin localization. However, the spin-gap at $\Gamma$ point ($k \, = \, 0,2\pi$) remains unchanged for any value of $\lambda$ with a fixed $U$ value (see Fig.2(b)). Moreover, the $\Delta_s(k)$ shows almost linear behavior around the Brillouin zone boundary ($k \, = \, \pi$) and its slope increases with increase in spin-orbit coupling strength. This is reminiscent of the behaviour of the edge-state bands in the non-interacting Kane-Mele model (see Fig.1(b)). Investigation of wider nanoribbons ($N_z=20$) shows similar qualitative features and trends. The most important feature which is seen for both narrow and wide nanoribbons is the complete vanishing of spin-gap at the Brillouin zone boundary, $k \, = \, \pi$ at all values of $\lambda$ and $U$. This is consistent with earlier reports based on multi-variable variational Monte Carlo (MVMC) calculations\citep{Yamaji}. However, the spin gap remains non-zero at all other $k$-points. It indicates that the gapless spin-excitation is unlikely for shorter length nanoribbons. To get better insight on this behavior, we further investigate the spin-ordering in the system over the full Brillouin zone.

\subsection{Spin density}

  \begin{figure}		
	\includegraphics[scale=0.32]{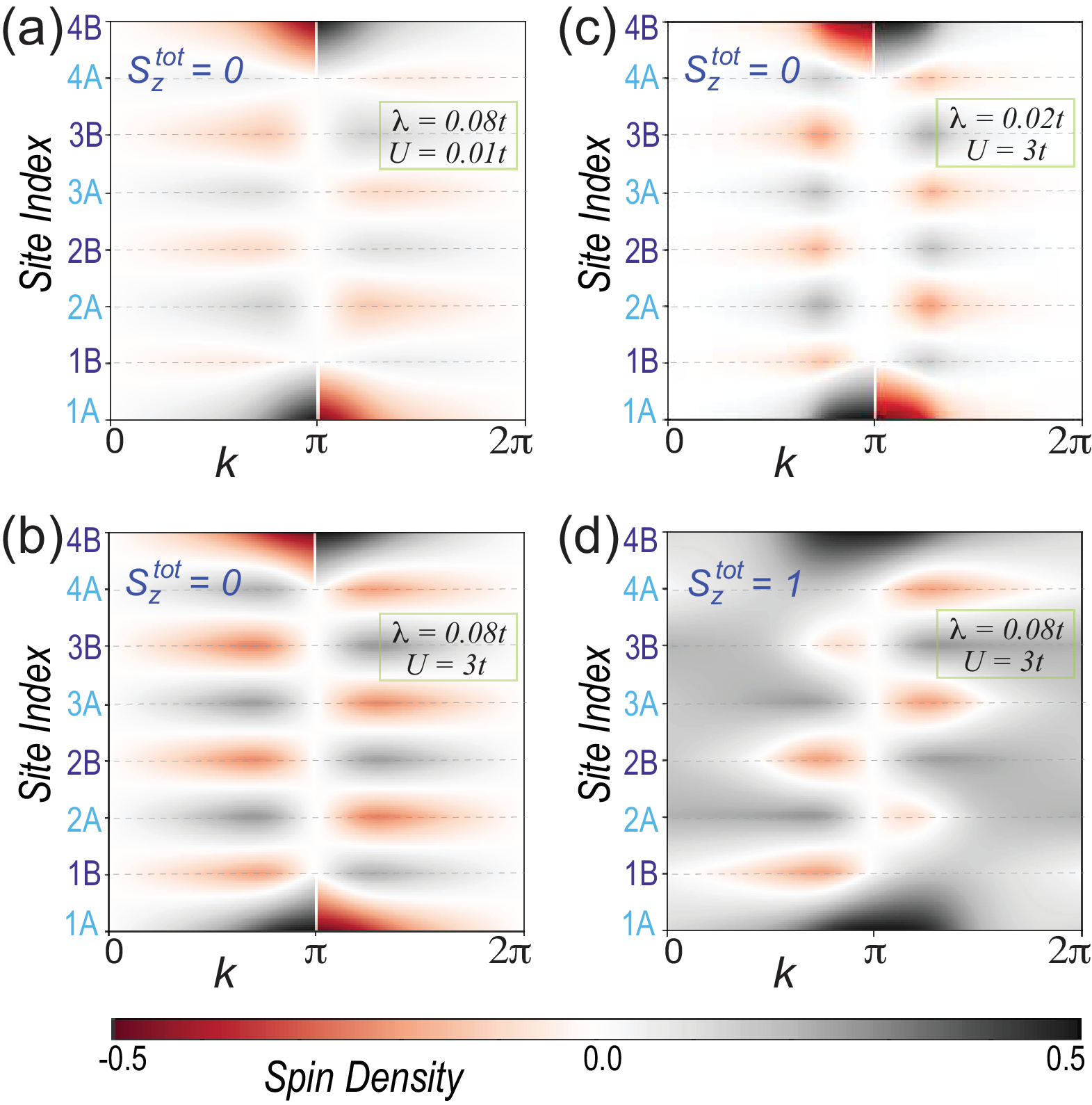}
	\caption{(Color online) The spin density profiles of a zigzag edge honeycomb nanoribbon with $N_z=4$, as a function of the momentum and site indices along the width of the ribbon. (a) and (b) show the spin density profiles for $U=0.01t$ and $U=3t$ in $S^{tot}_z=0$ spin sector, respectively for $\lambda=0,08t$. (c) show the same for $U=3t$ and $\lambda=0.02t$. (d) show the spin density profile in $S^{tot}_z=1$ spin sector for $U=3t$ and $\lambda=0.08t$. The color bar is provided with the spin density scale. The spin density profiles in the $S^{tot}_z=1$ spin sector for small $U$ values are not given due to very small magnitudes of spin density that is almost undectable in the color bar scale.}
  \end{figure}

We present the spin density profiles of half-filled zHNR with $N_z=4$ in both $S^{tot}_z=0$ and $S^{tot}_z=1$ spin sectors, with varying $\lambda$ and $U$ in Fig.3. We discuss the two spin sectors separately.

$\mathbf{S^{tot}_z=0}$: In absence of $\lambda$, the on-site Coulomb correlation results in edge spin-localization mainly at Brillouin zone boundary, i.e., at $k=\pi$ with negligible localization at any other parts of the Brillouin zone \cite{sdutta-scirep}. The bulk spin-polarization appears at higher $U$ value\cite{waka-ssp}. However, the spin density vanishes completely near the Brillouin zone boundary in presence of $\lambda$, as can be seen in Fig.3(a) and (b) for $\lambda=0.08t$. This in turn explains the vanishing spin gap at $k=\pi$. Since there is no preferred orientation of spin at $k=\pi$ in $S^{tot}_z=0$ state, a spin flip can occur at this momentum without incurring any energy cost. The spins at the opposite edges show antiferromagnetic correlation just next to $k=\pi$, due to nonzero $\Delta_s(k)$. Apart from the expected appearance of higher spin densities along the edge atoms, we find the appearance of a bulk antiferromagnetic spin-ordering as soon as the electronic interactions are turned on at very low value of $U=0.01t$ (see Fig.3(a)). This is in stark contrast with the case of zHNRs with only on-site Coulomb interaction where one does not find any bulk spin-polarization upto considerably large value of $U$ in absence of spin-orbit coupling \cite{sdutta-scirep}. The bulk spin-polarization further enhances with increase in on-site Coulomb correlation strength to $U/t=3$ (see Fig.3(b)), as expected. We observe that, the bulk spin-polarization is confined near the Dirac point for weaker spin-orbit coupling, as can be seen with $\lambda=0.02t$ (see Fig.3(c)) and with increase in $\lambda$, this gets extended over the other parts of the Brillouin zone. This can be attributed to the $\lambda$-induced helical spin-ordering in presence of even very small $U$. Another interesting feature is the change in the sign of the spin density at each site in the either sides of $k=\pi$. This along with $E(k)$ for the ground state provides information about how the many-body wavefunction is constructed from the single-particle eigenfunctions: the many-body wavefunction exhibits the nature of time-reversed partner states at $k$ points related by time-reversal symmetry ensuring $E(k=k_0)=E(k=-k_0+2\pi)$ where $k_0 \in [0,\pi]$. 

$\mathbf{S^{tot}_z=1}$ : Fig.3(d) shows the spin-density profile for $\lambda=0.08t$ with on-site Coulomb correlation $U=3t$ in the $S^{tot}_z=1$ spin sector. The edge atoms exhibit ferromagnetic correlation which is most prominent near $k=\pi$ while the bulk polarization is largely unaffected. This suggests that the spin flipping occurs at the edges resulting in an asymmetry in the spin-density profile of the system. To understand this, we can compare the spin-density at site $1A$ at $k>\pi$ for $S_z^{tot}=0$ and $S_z^{tot}=1$ with same values of $\lambda$ and $U$ (Figs.3(b) and (d)). Due to flipping of spin in $S_z^{tot}=1$ state, the density will change sign from negative to positive at site $1A$ and gets ferromagnetically coupled with the positive spin density at site $1B$. This causes the skewed profile of spin density near $k=\pi$ at site $1A$. Similar behaviour is also seen at the other edge. With gradual decrease in $U$, the bulk and edge spin-localization gradually decreases and gradual dominance of $\lambda$ induces spin canting. As a result, the spin density at the edges decreases gradually either with higher $\lambda$ or with smaller $U$ values.

The fact that the $S_z$ commutes with the Hamiltonian, allows us to obtain the spin density along the $z$ direction. However in the presence of spin-orbit coupling, the in-plane magnetization is favoured \citep{Rachel}. Thus, the competition between $U$ and $\lambda$ is expected to develop a helical spin ordering across the width of the nanoribbon as can be seen from the spin density plots. Our calculations with nanoribbons of width $N_z=20$ show similar features for both spin density and the spin gap.

\subsection{Many-body charge gap}

The signature of $Z_2$ topological order present in the Kane-Mele model is captured by the presence of robust edge states in nanoribbon systems.  Counter-intuitively, particle-hole symmetry of KM and KMH model ensures charge and spin-currents vanish on all bonds \citep{DongZheng11}. Hence, edge and charge spin currents are not good criteria to comment on the topological nature of the system in the presence of interactions. In Ref.\citep{Hohenadler11} and Ref.\citep{Hohenadler12}, the phase transition points have been identified by analysing single-particle gap in the $U-\lambda$ parameter space. The single-particle gap for the $2D$ system, evaluated at Dirac ($K$) point, shows a cusp at the critical points. This is possible because the onset of magnetic ordering breaks time-reversal symmetry and as a result the phase transition between TBI and AFMI phase can occur without the closing of single-particle gap. The question we now seek to answer is whether the signature of topological phase transition can be detected from an analysis of many-body charge gap profile of the nanoribbon over the full Brillouin zone, which is a more natural quantity, used to characterize the conduction properties of a system in the presence of dominant electron-electron interactions in reduced dimensions.

The many-body charge gap $(\Delta_c)$ is defined as:

\begin{equation}
    \Delta_c(k) \, = \, \left[E^{N+1}_0 \, + E^{N-1}_0 \, - \, 2E^N_0 \right](k)
\end{equation}

This can be understood as the energy difference between charging $\left(E^{N+1}_0 \, - \, E^N_0 \right)$ and discharging $\left(E^{N-1}_0 \, - \, E^N_0 \right)$ process in the ground state with $N$ number of electrons. Here $E^M_0(k)$ is the energy of the ground state wavefunction with $M$ electrons and $S_z^{\text{tot}} \, = \, 0 (0.5)$ for $\text{even}(\text{odd})$ $M$. To understand the effect of spin-orbit coupling on the charge-gap, we compute the $\Delta_c(k)$ for different values of  $U/t$ and $\lambda$. 

In Fig.4(a), we show the dispersion of $\Delta_c(k)$ for a fixed spin-orbit interaction strength ($\lambda=0.08t$) with varying Coulomb interaction term $U$. As can be seen, for very small Coulomb interaction strength ($U/t=0.2$), the nature of $\Delta_c(k)$ dispersion resembles the $\Delta_s(k)$ behavior with a minima at $k = \pi$, except non-vanishing charge gap at that momentum. With increase in $U/t$, the $\Delta_c(k)$ increases throughout the first Brillouin zone. Interestingly, the charge gap minima shifts from $k = \pi$ to the Dirac point with increase in $U/t$ value from $1$ to $2$. It should be noted that the $\Delta_c$ always attains its minima close to the Dirac point independent of the strength of the Hubbard interaction when the $\lambda$ is set to zero. This suggests a strong competition between the $U/t$ term and the $\lambda$ term in the Hamiltonian and the position of minimum $\Delta_{c}(k)$ is determined by the one which is dominant.

  \begin{figure}		
	\includegraphics[scale=0.32]{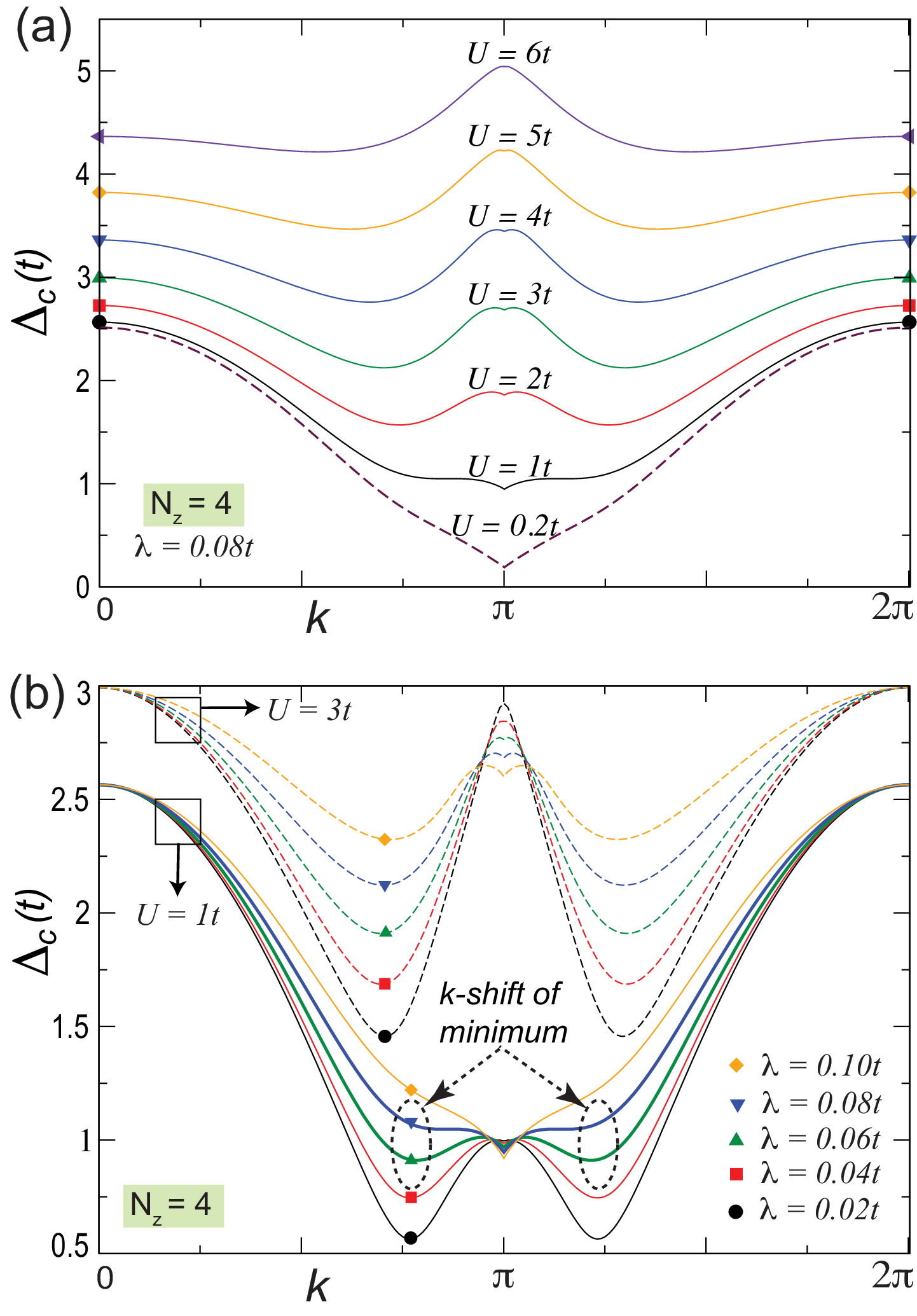}
	\caption{(Color online) The variation of charge gap ($(\Delta_c(k))$) as a function of momentum for a zigzag edge honeycomb nanoribbon with $N_z=4$ for (a) a fixed value of $\lambda=0.08t$ with $U/t$ varying from $0.2$ to $6$ and (b) two fixed values of $U/t=1$ and $U/t=3$ with $\lambda$ varying from $0.02t$ to $0.1t$. The shift of the charge gap minima from $k=\pi$ towards the Dirac point is marked with dotted elipses.}
  \end{figure}

To gain further insight in this, we plot the dispersions of $\Delta_c(k)$ with varying $\lambda$ for two fixed values of $U/t=1$ and $3$ in Fig.4(b). As can be seen, the charge gap at the $\Gamma$ point ($k=0,2\pi$) depends only on the value of $U/t$, irrespective of the spin-orbit coupling strength. For a fixed $U/t$ value, the charge gap increases with increase in $\lambda$, except near the Brillouin zone boundary ($k=\pi$) where it shows opposite trend. For $U/t=1$, the minima of $\Delta_{c}(k)$ appears at $k=\pi$ till the spin-orbit coupling strength of $0.08t$. Further decrease in the strength of $\lambda$ makes the $\Delta_{c}(k)$ minima shift towards the Dirac point. This is marked with dotted oval in Fig.4(b). This behavior further confirms the competition between $U/t$ and $\lambda$ to determine the location of charge gap minima in the momentum space. When the strength of $U/t$ is further raised to $3$, the minima of $\Delta_{c}(k)$ always appears near the Dirac point, irrespective of the strength of $\lambda$, considered here.

The above observations indicate that, whenever the spin-orbit coupling is dominant, the charge gap attains its minima at $k=\pi$. On the other hand, the dominance of Coulomb correlation shifts the charge gap minima towards the Dirac point. We interpret this as a signature of phase transition in the momentum space from the TBI phase to the AFMI phase. This can be characterized by the spin and charge gap behaviors that we observe here. We observe gapless spin excitations for $\lambda$ $\in$ $[0.02t,0.10t]$ and $U/t$ $\in$ $[1,6]$. However, the charge gap analysis indicates the existence two phases, i.e., TBI phase when $\lambda$ dominates and AFMI when $U$ dominates. Our results indicate the phase transition from TBI to AFMI phase when $\lambda$ is decreased from $0.08t$ to $0.06t$ for $U/t=1$. Beyond the Coulomb correlation strength of $U/t=1$, we observe AFMI phase, irrespective of the strength of the spin-orbit coupling. Our results for the phases of KMH Hamiltonian defined on zHNRs, match with the phases predicted in Ref.\citep{Yamaji}. The quantitative differences of the phase boundaries in terms of the strength of the parameters, $U/t$ and $\lambda$ can be attributed to the differences of the numerical techniques and associated approximations therein. Nevertheless, our study provides a new way to characterize and identify the quantum phases of KMH Hamiltonian from the dispersion behavior of spin and charge gaps in the momentum space.

Since, most of the earlier studies on Kane-Mele Hubbard model were restricted to analysis of various quantities at special points in Brillouin zone ($\Gamma$ and $\mathbf{K}$)(\citep{Hohenadler11},\citep{Hohenadler12}), the momentum shift of the charge-gap minima has not been reported before within many-body formalism. However, similar signature has been observed within previous mean-field studies\citep{Soriano,Lado}. Actually, when $\lambda$ is dominant over the $U/t$, the mean-field band-structure resembles with that of the non-interacting Kane-Mele model\citep{kane-mele1}. Mean-field band structure exhibits minima of band gap at $k=\pi$ and the non-interacting band structure shows gap closing at the same point. Note that, we observe similar dispersions of spin gap for all $U/t$ and $\lambda$ values (see Fig.2). The charge gap too exhibits similar dispersion for $U/t=0.2$ and $\lambda=0.08t$, only with a non-zero gap at $k=\pi$ (see Fig.4(a)). This indicates that the system is still in TBI phase. However, as one decreases the spin-orbit coupling strength for a fixed onsite Coulomb correlation, the point at which the mean-field gap closes, shifts away from the Brillouin zone boundary towards the Dirac point before eventually opening up\citep{Soriano}. This has been classified as a transition from topological insulating phase to a valley half-metal with antiferromagnetically ordered spins at the edges, that can be understood via perturbative arguments \citep{kane-mele1,Soriano}. This shifting of the gap-closing point at mean-field level is nothing but a signature of the shifting of the charge-gap minima that we observed in the many-body CI calculations. However, instead of marking a transition from TBI phase to a valley half-metal, it actually marks a transition from a TBI phase to AFMI phase, consistent with the results of Ref.\citep{Yamaji}. Thus, results from our CI calculations bridges the general understanding of the available mean-field and many-body results for KMH model.

\section{\label{sec:level4} Conclusions \protect}

Here we study the Kane-Mele-Hubbard model by investigating the magnetic and conduction properties of zigzag-edged honeycomb nanoribbons via many-body CI method. We report the exact many-body charge-gap and spin-gap dispersion profiles over the full first Brillouin zone. Our exact results for narrow nanoribbons ($N_z=4$) indicate that the spin-sector retains the topological nature and remains gapless at all values of spin-orbit coupling and Hubbard correlation strength. On the other hand, the charge-sector undergoes a transition from the TBI phase to AFMI phase as one increases the strength of Hubbard interaction. This is marked by a shift of the charge-gap minima from the Brillouin zone boundary to the Dirac point. At mean-field level with off-plane spin alignment, this manifests itself as shifting of the gap-closing point from the Brillouin-zone boundary to the Dirac point. Further investigations for wider nanoribbons of width $N_z=20$ by considering complete active space near the Fermi level shows similar qualitative behaviors. Our study provides a new way to identify the phase transitions of KMH model in terms of the changes in the dispersions of many-body charge and spin gaps in case of zHNRs.  

\section{\label{sec:level5}Acknowledgement \protect}
SM acknowledges the SEED grant (RGEMS) no. SG20230062, funded by Vellore Institute of Technology (VIT), Vellore, India. KW acknowledges the JSPS KAKENHI (Nos. 22H05473, JP21H01019, JP18H01154) and JST CREST (No. JPMJCR19T1) research grants. SD thanks IISER Tirupati for Intramural Funding and SERB, Department of Science and Technology (DST), Govt. of India for research grant CRG/2021/001731. SD and KW acknowledge the support from Hyogo Overseas Research Network (HORN) project.

%\bibliography{apssamp.bib}% Produces the bibliography via BibTeX.

\begin{thebibliography}{34}%
\makeatletter
\providecommand \@ifxundefined [1]{%
 \@ifx{#1\undefined}
}%
\providecommand \@ifnum [1]{%
 \ifnum #1\expandafter \@firstoftwo
 \else \expandafter \@secondoftwo
 \fi
}%
\providecommand \@ifx [1]{%
 \ifx #1\expandafter \@firstoftwo
 \else \expandafter \@secondoftwo
 \fi
}%
\providecommand \natexlab [1]{#1}%
\providecommand \enquote  [1]{``#1''}%
\providecommand \bibnamefont  [1]{#1}%
\providecommand \bibfnamefont [1]{#1}%
\providecommand \citenamefont [1]{#1}%
\providecommand \href@noop [0]{\@secondoftwo}%
\providecommand \href [0]{\begingroup \@sanitize@url \@href}%
\providecommand \@href[1]{\@@startlink{#1}\@@href}%
\providecommand \@@href[1]{\endgroup#1\@@endlink}%
\providecommand \@sanitize@url [0]{\catcode `\\12\catcode `\$12\catcode
  `\&12\catcode `\#12\catcode `\^12\catcode `\_12\catcode `\%12\relax}%
\providecommand \@@startlink[1]{}%
\providecommand \@@endlink[0]{}%
\providecommand \url  [0]{\begingroup\@sanitize@url \@url }%
\providecommand \@url [1]{\endgroup\@href {#1}{\urlprefix }}%
\providecommand \urlprefix  [0]{URL }%
\providecommand \Eprint [0]{\href }%
\providecommand \doibase [0]{https://doi.org/}%
\providecommand \selectlanguage [0]{\@gobble}%
\providecommand \bibinfo  [0]{\@secondoftwo}%
\providecommand \bibfield  [0]{\@secondoftwo}%
\providecommand \translation [1]{[#1]}%
\providecommand \BibitemOpen [0]{}%
\providecommand \bibitemStop [0]{}%
\providecommand \bibitemNoStop [0]{.\EOS\space}%
\providecommand \EOS [0]{\spacefactor3000\relax}%
\providecommand \BibitemShut  [1]{\csname bibitem#1\endcsname}%
\let\auto@bib@innerbib\@empty
%</preamble>
\bibitem [{\citenamefont {Kane}\ and\ \citenamefont
  {Mele}(2005{\natexlab{a}})}]{kane-mele1}%
  \BibitemOpen
  \bibfield  {author} {\bibinfo {author} {\bibfnamefont {C.~L.}\ \bibnamefont
  {Kane}}\ and\ \bibinfo {author} {\bibfnamefont {E.~J.}\ \bibnamefont
  {Mele}},\ }\href@noop {} {\bibfield  {journal} {\bibinfo  {journal} {Phys.
  Rev. Lett.}\ }\textbf {\bibinfo {volume} {95}},\ \bibinfo {pages} {226801}
  (\bibinfo {year} {2005}{\natexlab{a}})}\BibitemShut {NoStop}%
\bibitem [{\citenamefont {Haldane}(1988)}]{haldane}%
  \BibitemOpen
  \bibfield  {author} {\bibinfo {author} {\bibfnamefont {F.~D.~M.}\
  \bibnamefont {Haldane}},\ }\href@noop {} {\bibfield  {journal} {\bibinfo
  {journal} {Phys. Rev. Lett.}\ }\textbf {\bibinfo {volume} {61}},\ \bibinfo
  {pages} {2015} (\bibinfo {year} {1988})}\BibitemShut {NoStop}%
\bibitem [{\citenamefont {Moore}(2010)}]{moore}%
  \BibitemOpen
  \bibfield  {author} {\bibinfo {author} {\bibfnamefont {J.}~\bibnamefont
  {Moore}},\ }\href@noop {} {\bibfield  {journal} {\bibinfo  {journal}
  {Nature}\ }\textbf {\bibinfo {volume} {464}},\ \bibinfo {pages} {194}
  (\bibinfo {year} {2010})}\BibitemShut {NoStop}%
\bibitem [{\citenamefont {Murakami}\ \emph {et~al.}(2004)\citenamefont
  {Murakami}, \citenamefont {Nagaosa},\ and\ \citenamefont {Zhang}}]{murakami}%
  \BibitemOpen
  \bibfield  {author} {\bibinfo {author} {\bibfnamefont {S.}~\bibnamefont
  {Murakami}}, \bibinfo {author} {\bibfnamefont {N.}~\bibnamefont {Nagaosa}},\
  and\ \bibinfo {author} {\bibfnamefont {S.~C.}\ \bibnamefont {Zhang}},\
  }\href@noop {} {\bibfield  {journal} {\bibinfo  {journal} {Phys. Rev. Lett.}\
  }\textbf {\bibinfo {volume} {93}},\ \bibinfo {pages} {156804} (\bibinfo
  {year} {2004})}\BibitemShut {NoStop}%
\bibitem [{\citenamefont {Bernevig}\ \emph {et~al.}(2006)\citenamefont
  {Bernevig}, \citenamefont {Hughes},\ and\ \citenamefont {Zhang}}]{bernevig}%
  \BibitemOpen
  \bibfield  {author} {\bibinfo {author} {\bibfnamefont {B.~A.}\ \bibnamefont
  {Bernevig}}, \bibinfo {author} {\bibfnamefont {T.~L.}\ \bibnamefont
  {Hughes}},\ and\ \bibinfo {author} {\bibfnamefont {S.~C.}\ \bibnamefont
  {Zhang}},\ }\href@noop {} {\bibfield  {journal} {\bibinfo  {journal}
  {Science}\ }\textbf {\bibinfo {volume} {314}},\ \bibinfo {pages} {1757}
  (\bibinfo {year} {2006})}\BibitemShut {NoStop}%
\bibitem [{\citenamefont {Konig}\ \emph {et~al.}(2007)\citenamefont {Konig},
  \citenamefont {Wiedmann}, \citenamefont {Brune}, \citenamefont {Roth},
  \citenamefont {Buhmann}, \citenamefont {Molenkamp}, \citenamefont {Qi},\ and\
  \citenamefont {Zhang}}]{konig}%
  \BibitemOpen
  \bibfield  {author} {\bibinfo {author} {\bibfnamefont {M.}~\bibnamefont
  {Konig}}, \bibinfo {author} {\bibfnamefont {S.}~\bibnamefont {Wiedmann}},
  \bibinfo {author} {\bibfnamefont {C.}~\bibnamefont {Brune}}, \bibinfo
  {author} {\bibfnamefont {A.}~\bibnamefont {Roth}}, \bibinfo {author}
  {\bibfnamefont {H.}~\bibnamefont {Buhmann}}, \bibinfo {author} {\bibfnamefont
  {L.~W.}\ \bibnamefont {Molenkamp}}, \bibinfo {author} {\bibfnamefont {X.~L.}\
  \bibnamefont {Qi}},\ and\ \bibinfo {author} {\bibfnamefont {S.~C.}\
  \bibnamefont {Zhang}},\ }\href@noop {} {\bibfield  {journal} {\bibinfo
  {journal} {Science}\ }\textbf {\bibinfo {volume} {318}},\ \bibinfo {pages}
  {766} (\bibinfo {year} {2007})}\BibitemShut {NoStop}%
\bibitem [{\citenamefont {Kane}\ and\ \citenamefont
  {Mele}(2005{\natexlab{b}})}]{kane-mele2}%
  \BibitemOpen
  \bibfield  {author} {\bibinfo {author} {\bibfnamefont {C.~L.}\ \bibnamefont
  {Kane}}\ and\ \bibinfo {author} {\bibfnamefont {E.~J.}\ \bibnamefont
  {Mele}},\ }\href@noop {} {\bibfield  {journal} {\bibinfo  {journal} {Phys.
  Rev. Lett.}\ }\textbf {\bibinfo {volume} {95}},\ \bibinfo {pages} {146802}
  (\bibinfo {year} {2005}{\natexlab{b}})}\BibitemShut {NoStop}%
\bibitem [{\citenamefont {Roy}(2009)}]{roy}%
  \BibitemOpen
  \bibfield  {author} {\bibinfo {author} {\bibfnamefont {R.}~\bibnamefont
  {Roy}},\ }\href@noop {} {\bibfield  {journal} {\bibinfo  {journal} {Phys.
  Rev. B}\ }\textbf {\bibinfo {volume} {79}},\ \bibinfo {pages} {195321}
  (\bibinfo {year} {2009})}\BibitemShut {NoStop}%
\bibitem [{\citenamefont {Shitade}\ \emph {et~al.}(2009)\citenamefont
  {Shitade}, \citenamefont {Katsura}, \citenamefont {Kunes}, \citenamefont
  {Qi}, \citenamefont {Zhang},\ and\ \citenamefont {Nagaosa}}]{shitade}%
  \BibitemOpen
  \bibfield  {author} {\bibinfo {author} {\bibfnamefont {A.}~\bibnamefont
  {Shitade}}, \bibinfo {author} {\bibfnamefont {H.}~\bibnamefont {Katsura}},
  \bibinfo {author} {\bibfnamefont {J.}~\bibnamefont {Kunes}}, \bibinfo
  {author} {\bibfnamefont {X.~L.}\ \bibnamefont {Qi}}, \bibinfo {author}
  {\bibfnamefont {S.~C.}\ \bibnamefont {Zhang}},\ and\ \bibinfo {author}
  {\bibfnamefont {N.}~\bibnamefont {Nagaosa}},\ }\href@noop {} {\bibfield
  {journal} {\bibinfo  {journal} {Phys. Rev. Lett.}\ }\textbf {\bibinfo
  {volume} {102}},\ \bibinfo {pages} {256403} (\bibinfo {year}
  {2009})}\BibitemShut {NoStop}%
\bibitem [{\citenamefont {Laubach}\ \emph {et~al.}(2017)\citenamefont
  {Laubach}, \citenamefont {Reuther}, \citenamefont {Thomale},\ and\
  \citenamefont {Rachel}}]{laubach}%
  \BibitemOpen
  \bibfield  {author} {\bibinfo {author} {\bibfnamefont {M.}~\bibnamefont
  {Laubach}}, \bibinfo {author} {\bibfnamefont {J.}~\bibnamefont {Reuther}},
  \bibinfo {author} {\bibfnamefont {R.}~\bibnamefont {Thomale}},\ and\ \bibinfo
  {author} {\bibfnamefont {S.}~\bibnamefont {Rachel}},\ }\href@noop {}
  {\bibfield  {journal} {\bibinfo  {journal} {Phys. Rev. B}\ }\textbf {\bibinfo
  {volume} {96}},\ \bibinfo {pages} {121110(R)} (\bibinfo {year}
  {2017})}\BibitemShut {NoStop}%
\bibitem [{\citenamefont {Kim}\ \emph {et~al.}(2018)\citenamefont {Kim},
  \citenamefont {Kim}, \citenamefont {Sandilands}, \citenamefont {Sinn},
  \citenamefont {Lee}, \citenamefont {Son}, \citenamefont {Lee}, \citenamefont
  {Choi}, \citenamefont {Kim}, \citenamefont {Park}, \citenamefont {Jeon},
  \citenamefont {Kim}, \citenamefont {Park}, \citenamefont {Park},
  \citenamefont {Moon},\ and\ \citenamefont {Noh}}]{sykim}%
  \BibitemOpen
  \bibfield  {author} {\bibinfo {author} {\bibfnamefont {S.~Y.}\ \bibnamefont
  {Kim}}, \bibinfo {author} {\bibfnamefont {T.~Y.}\ \bibnamefont {Kim}},
  \bibinfo {author} {\bibfnamefont {L.~J.}\ \bibnamefont {Sandilands}},
  \bibinfo {author} {\bibfnamefont {S.}~\bibnamefont {Sinn}}, \bibinfo {author}
  {\bibfnamefont {M.~C.}\ \bibnamefont {Lee}}, \bibinfo {author} {\bibfnamefont
  {J.}~\bibnamefont {Son}}, \bibinfo {author} {\bibfnamefont {S.}~\bibnamefont
  {Lee}}, \bibinfo {author} {\bibfnamefont {K.~Y.}\ \bibnamefont {Choi}},
  \bibinfo {author} {\bibfnamefont {W.}~\bibnamefont {Kim}}, \bibinfo {author}
  {\bibfnamefont {B.~G.}\ \bibnamefont {Park}}, \bibinfo {author}
  {\bibfnamefont {C.}~\bibnamefont {Jeon}}, \bibinfo {author} {\bibfnamefont
  {H.~D.}\ \bibnamefont {Kim}}, \bibinfo {author} {\bibfnamefont {C.~H.}\
  \bibnamefont {Park}}, \bibinfo {author} {\bibfnamefont {J.~G.}\ \bibnamefont
  {Park}}, \bibinfo {author} {\bibfnamefont {S.~J.}~\bibnamefont {Moon}},\ and\
  \bibinfo {author} {\bibfnamefont {T.~W.}~\bibnamefont {Noh}},\ }\href@noop {}
  {\bibfield  {journal} {\bibinfo  {journal} {Phys. Rev. Lett.}\ }\textbf
  {\bibinfo {volume} {120}},\ \bibinfo {pages} {136402} (\bibinfo {year}
  {2018})}\BibitemShut {NoStop}%
\bibitem [{\citenamefont {Gong}\ \emph {et~al.}(2017)\citenamefont {Gong},
  \citenamefont {Li}, \citenamefont {Li}, \citenamefont {Ji}, \citenamefont
  {Stern}, \citenamefont {Xia}, \citenamefont {Cao}, \citenamefont {Bao},
  \citenamefont {Wang}, \citenamefont {Wang}, \citenamefont {Qiu},
  \citenamefont {Cava}, \citenamefont {Louie}, \citenamefont {Xia},\ and\
  \citenamefont {Zhang}}]{chenggong}%
  \BibitemOpen
  \bibfield  {author} {\bibinfo {author} {\bibfnamefont {C.}~\bibnamefont
  {Gong}}, \bibinfo {author} {\bibfnamefont {L.}~\bibnamefont {Li}}, \bibinfo
  {author} {\bibfnamefont {Z.}~\bibnamefont {Li}}, \bibinfo {author}
  {\bibfnamefont {H.}~\bibnamefont {Ji}}, \bibinfo {author} {\bibfnamefont
  {A.}~\bibnamefont {Stern}}, \bibinfo {author} {\bibfnamefont
  {Y.}~\bibnamefont {Xia}}, \bibinfo {author} {\bibfnamefont {T.}~\bibnamefont
  {Cao}}, \bibinfo {author} {\bibfnamefont {W.}~\bibnamefont {Bao}}, \bibinfo
  {author} {\bibfnamefont {C.}~\bibnamefont {Wang}}, \bibinfo {author}
  {\bibfnamefont {Y.}~\bibnamefont {Wang}}, \bibinfo {author} {\bibfnamefont
  {Z.~Q.}\ \bibnamefont {Qiu}}, \bibinfo {author} {\bibfnamefont {R.~J.}\
  \bibnamefont {Cava}}, \bibinfo {author} {\bibfnamefont {S.~G.}\ \bibnamefont
  {Louie}}, \bibinfo {author} {\bibfnamefont {J.}~\bibnamefont {Xia}},\ and\
  \bibinfo {author} {\bibfnamefont {X.}~\bibnamefont {Zhang}},\ }\href@noop {}
  {\bibfield  {journal} {\bibinfo  {journal} {Nature}\ }\textbf {\bibinfo
  {volume} {546}},\ \bibinfo {pages} {265} (\bibinfo {year}
  {2017})}\BibitemShut {NoStop}%
\bibitem [{\citenamefont {Lee}\ \emph {et~al.}(2016)\citenamefont {Lee},
  \citenamefont {Lee}, \citenamefont {Ryoo}, \citenamefont {Kang},
  \citenamefont {Kim}, \citenamefont {Kim}, \citenamefont {Park}, \citenamefont
  {Park},\ and\ \citenamefont {Cheong}}]{jaeunglee}%
  \BibitemOpen
  \bibfield  {author} {\bibinfo {author} {\bibfnamefont {J.~U.}\ \bibnamefont
  {Lee}}, \bibinfo {author} {\bibfnamefont {S.}~\bibnamefont {Lee}}, \bibinfo
  {author} {\bibfnamefont {J.~H.}\ \bibnamefont {Ryoo}}, \bibinfo {author}
  {\bibfnamefont {S.}~\bibnamefont {Kang}}, \bibinfo {author} {\bibfnamefont
  {T.~Y.}\ \bibnamefont {Kim}}, \bibinfo {author} {\bibfnamefont
  {P.}~\bibnamefont {Kim}}, \bibinfo {author} {\bibfnamefont {C.~H.}\
  \bibnamefont {Park}}, \bibinfo {author} {\bibfnamefont {J.~G.}\ \bibnamefont
  {Park}},\ and\ \bibinfo {author} {\bibfnamefont {H.}~\bibnamefont {Cheong}},\
  }\href@noop {} {\bibfield  {journal} {\bibinfo  {journal} {Nano Lett.}\
  }\textbf {\bibinfo {volume} {16}},\ \bibinfo {pages} {7433} (\bibinfo {year}
  {2016})}\BibitemShut {NoStop}%
\bibitem [{\citenamefont {Kuo}\ \emph {et~al.}(2016)\citenamefont {Kuo},
  \citenamefont {Neumann}, \citenamefont {Balamurugan}, \citenamefont {Park},
  \citenamefont {Kang}, \citenamefont {Shiu}, \citenamefont {Kang},
  \citenamefont {Hong}, \citenamefont {Han}, \citenamefont {Noh},\ and\
  \citenamefont {Park}}]{chengtaikuo}%
  \BibitemOpen
  \bibfield  {author} {\bibinfo {author} {\bibfnamefont {C.~T.}\ \bibnamefont
  {Kuo}}, \bibinfo {author} {\bibfnamefont {M.}~\bibnamefont {Neumann}},
  \bibinfo {author} {\bibfnamefont {K.}~\bibnamefont {Balamurugan}}, \bibinfo
  {author} {\bibfnamefont {H.~J.}\ \bibnamefont {Park}}, \bibinfo {author}
  {\bibfnamefont {S.}~\bibnamefont {Kang}}, \bibinfo {author} {\bibfnamefont
  {H.~W.}\ \bibnamefont {Shiu}}, \bibinfo {author} {\bibfnamefont {J.~H.}\
  \bibnamefont {Kang}}, \bibinfo {author} {\bibfnamefont {B.~H.}\ \bibnamefont
  {Hong}}, \bibinfo {author} {\bibfnamefont {M.}~\bibnamefont {Han}}, \bibinfo
  {author} {\bibfnamefont {T.~W.}\ \bibnamefont {Noh}},\ and\ \bibinfo {author}
  {\bibfnamefont {J.~G.}\ \bibnamefont {Park}},\ }\href@noop {} {\bibfield
  {journal} {\bibinfo  {journal} {Scientific Reports}\ }\textbf {\bibinfo
  {volume} {6}},\ \bibinfo {pages} {20904} (\bibinfo {year}
  {2016})}\BibitemShut {NoStop}%
\bibitem [{\citenamefont {Mishra}\ and\ \citenamefont {Lee}(2018)}]{archana}%
  \BibitemOpen
  \bibfield  {author} {\bibinfo {author} {\bibfnamefont {A.}~\bibnamefont
  {Mishra}}\ and\ \bibinfo {author} {\bibfnamefont {S.~B.}\ \bibnamefont
  {Lee}},\ }\href@noop {} {\bibfield  {journal} {\bibinfo  {journal}
  {Scientific Reports}\ }\textbf {\bibinfo {volume} {8}},\ \bibinfo {pages}
  {799} (\bibinfo {year} {2018})}\BibitemShut {NoStop}%
\bibitem [{\citenamefont {Mishra}\ and\ \citenamefont {Lee}(2019)}]{mishra}%
  \BibitemOpen
  \bibfield  {author} {\bibinfo {author} {\bibfnamefont {A.}~\bibnamefont
  {Mishra}}\ and\ \bibinfo {author} {\bibfnamefont {S.~B.}\ \bibnamefont
  {Lee}},\ }\href@noop {} {\bibfield  {journal} {\bibinfo  {journal} {Phys.
  Rev. B}\ }\textbf {\bibinfo {volume} {100}},\ \bibinfo {pages} {045146}
  (\bibinfo {year} {2019})}\BibitemShut {NoStop}%
\bibitem [{\citenamefont {Dutta}\ \emph {et~al.}(2008)\citenamefont {Dutta},
  \citenamefont {Lakshmi},\ and\ \citenamefont {Pati}}]{sdutta-physrev}%
  \BibitemOpen
  \bibfield  {author} {\bibinfo {author} {\bibfnamefont {S.}~\bibnamefont
  {Dutta}}, \bibinfo {author} {\bibfnamefont {S.}~\bibnamefont {Lakshmi}},\
  and\ \bibinfo {author} {\bibfnamefont {S.~K.}\ \bibnamefont {Pati}},\
  }\href@noop {} {\bibfield  {journal} {\bibinfo  {journal} {Phys. Rev. B}\
  }\textbf {\bibinfo {volume} {77}},\ \bibinfo {pages} {073412} (\bibinfo
  {year} {2008})}\BibitemShut {NoStop}%
\bibitem [{\citenamefont {Dutta}\ and\ \citenamefont
  {Wakabayashi}(2012)}]{sdutta-scirep}%
  \BibitemOpen
  \bibfield  {author} {\bibinfo {author} {\bibfnamefont {S.}~\bibnamefont
  {Dutta}}\ and\ \bibinfo {author} {\bibfnamefont {K.}~\bibnamefont
  {Wakabayashi}},\ }\href@noop {} {\bibfield  {journal} {\bibinfo  {journal}
  {Scientific Reports}\ }\textbf {\bibinfo {volume} {2}},\ \bibinfo {pages}
  {00519} (\bibinfo {year} {2012})}\BibitemShut {NoStop}%
\bibitem [{\citenamefont {Fujita}\ \emph {et~al.}(1996)\citenamefont {Fujita},
  \citenamefont {Wakabayashi}, \citenamefont {Nakada},\ and\ \citenamefont
  {Kusakabe}}]{waka1}%
  \BibitemOpen
  \bibfield  {author} {\bibinfo {author} {\bibfnamefont {M.}~\bibnamefont
  {Fujita}}, \bibinfo {author} {\bibfnamefont {K.}~\bibnamefont {Wakabayashi}},
  \bibinfo {author} {\bibfnamefont {K.}~\bibnamefont {Nakada}},\ and\ \bibinfo
  {author} {\bibfnamefont {K.}~\bibnamefont {Kusakabe}},\ }\href@noop {}
  {\bibfield  {journal} {\bibinfo  {journal} {J. Phys. Soc. Jpn.}\ }\textbf
  {\bibinfo {volume} {65}},\ \bibinfo {pages} {1920} (\bibinfo {year}
  {1996})}\BibitemShut {NoStop}%
\bibitem [{\citenamefont {Liu}\ and\ \citenamefont
  {Wakabayashi}(2017)}]{waka2}%
  \BibitemOpen
  \bibfield  {author} {\bibinfo {author} {\bibfnamefont {F.}~\bibnamefont
  {Liu}}\ and\ \bibinfo {author} {\bibfnamefont {K.}~\bibnamefont
  {Wakabayashi}},\ }\href@noop {} {\bibfield  {journal} {\bibinfo  {journal}
  {Phys. Rev. Lett.}\ }\textbf {\bibinfo {volume} {118}},\ \bibinfo {pages}
  {076803} (\bibinfo {year} {2017})}\BibitemShut {NoStop}%
\bibitem [{\citenamefont {Liu}\ \emph {et~al.}(2017)\citenamefont {Liu},
  \citenamefont {Yamamoto},\ and\ \citenamefont {Wakabayashi}}]{waka3}%
  \BibitemOpen
  \bibfield  {author} {\bibinfo {author} {\bibfnamefont {F.}~\bibnamefont
  {Liu}}, \bibinfo {author} {\bibfnamefont {M.}~\bibnamefont {Yamamoto}},\ and\
  \bibinfo {author} {\bibfnamefont {K.}~\bibnamefont {Wakabayashi}},\
  }\href@noop {} {\bibfield  {journal} {\bibinfo  {journal} {J. Phys. Soc.
  Jpn.}\ }\textbf {\bibinfo {volume} {86}},\ \bibinfo {pages} {123707}
  (\bibinfo {year} {2017})}\BibitemShut {NoStop}%
\bibitem [{\citenamefont {Zarea}\ and\ \citenamefont {Sandler}(2009)}]{zarea}%
  \BibitemOpen
  \bibfield  {author} {\bibinfo {author} {\bibfnamefont {M.}~\bibnamefont
  {Zarea}}\ and\ \bibinfo {author} {\bibfnamefont {N.}~\bibnamefont
  {Sandler}},\ }\href@noop {} {\bibfield  {journal} {\bibinfo  {journal}
  {Physica B: Condensed Matter}\ }\textbf {\bibinfo {volume} {404}},\ \bibinfo
  {pages} {2694} (\bibinfo {year} {2009})}\BibitemShut {NoStop}%
\bibitem [{\citenamefont {Wakabayashi}\ and\ \citenamefont
  {Dutta}(2012)}]{waka-ssp}%
  \BibitemOpen
  \bibfield  {author} {\bibinfo {author} {\bibfnamefont {K.}~\bibnamefont
  {Wakabayashi}}\ and\ \bibinfo {author} {\bibfnamefont {S.}~\bibnamefont
  {Dutta}},\ }\href@noop {} {\bibfield  {journal} {\bibinfo  {journal} {Solid
  State Comm.}\ }\textbf {\bibinfo {volume} {152}},\ \bibinfo {pages} {1420}
  (\bibinfo {year} {2012})}\BibitemShut {NoStop}%
\bibitem [{\citenamefont {Chung}\ \emph {et~al.}(2014)\citenamefont {Chung},
  \citenamefont {Lee},\ and\ \citenamefont {Chao}}]{chung}%
  \BibitemOpen
  \bibfield  {author} {\bibinfo {author} {\bibfnamefont {C.~H.}\ \bibnamefont
  {Chung}}, \bibinfo {author} {\bibfnamefont {D.~H.}\ \bibnamefont {Lee}},\
  and\ \bibinfo {author} {\bibfnamefont {S.~P.}\ \bibnamefont {Chao}},\
  }\href@noop {} {\bibfield  {journal} {\bibinfo  {journal} {Phys. Rev. B}\
  }\textbf {\bibinfo {volume} {90}},\ \bibinfo {pages} {035116} (\bibinfo
  {year} {2014})}\BibitemShut {NoStop}%
\bibitem [{\citenamefont {Makarova}\ \emph {et~al.}(2015)\citenamefont
  {Makarova}, \citenamefont {Shelankov}, \citenamefont {Zyrianova},
  \citenamefont {Veinger}, \citenamefont {Tisnek}, \citenamefont {Lähderanta},
  \citenamefont {Shames}, \citenamefont {Okotrub}, \citenamefont {Bulusheva},
  \citenamefont {Chekhova}, \citenamefont {Pinakov}, \citenamefont {Asanov},\
  and\ \citenamefont {Sljivancanin}}]{makarova-scirep}%
  \BibitemOpen
  \bibfield  {author} {\bibinfo {author} {\bibfnamefont {T.~L.}\ \bibnamefont
  {Makarova}}, \bibinfo {author} {\bibfnamefont {A.~L.}\ \bibnamefont
  {Shelankov}}, \bibinfo {author} {\bibfnamefont {A.~A.}\ \bibnamefont
  {Zyrianova}}, \bibinfo {author} {\bibfnamefont {A.~I.}\ \bibnamefont
  {Veinger}}, \bibinfo {author} {\bibfnamefont {T.~V.}\ \bibnamefont {Tisnek}},
  \bibinfo {author} {\bibfnamefont {E.}~\bibnamefont {Lähderanta}}, \bibinfo
  {author} {\bibfnamefont {A.~I.}\ \bibnamefont {Shames}}, \bibinfo {author}
  {\bibfnamefont {A.~V.}\ \bibnamefont {Okotrub}}, \bibinfo {author}
  {\bibfnamefont {L.~G.}\ \bibnamefont {Bulusheva}}, \bibinfo {author}
  {\bibfnamefont {G.~N.}\ \bibnamefont {Chekhova}}, \bibinfo {author}
  {\bibfnamefont {D.~V.}\ \bibnamefont {Pinakov}}, \bibinfo {author}
  {\bibfnamefont {I.~P.}\ \bibnamefont {Asanov}},\ and\ \bibinfo {author}
  {\bibfnamefont {Z.}~\bibnamefont {Sljivancanin}},\ }\href@noop {} {\bibfield
  {journal} {\bibinfo  {journal} {Sci. Rep.}\ }\textbf {\bibinfo {volume}
  {5}},\ \bibinfo {pages} {13382} (\bibinfo {year} {2015})}\BibitemShut
  {NoStop}%
\bibitem [{\citenamefont {Baringhaus}\ \emph {et~al.}(2013)\citenamefont
  {Baringhaus}, \citenamefont {Edler},\ and\ \citenamefont
  {Tegenkamp}}]{baringhaus-jpcm}%
  \BibitemOpen
  \bibfield  {author} {\bibinfo {author} {\bibfnamefont {J.}~\bibnamefont
  {Baringhaus}}, \bibinfo {author} {\bibfnamefont {F.}~\bibnamefont {Edler}},\
  and\ \bibinfo {author} {\bibfnamefont {C.}~\bibnamefont {Tegenkamp}},\
  }\href@noop {} {\bibfield  {journal} {\bibinfo  {journal} {J. Phys.: Condens.
  Matter}\ }\textbf {\bibinfo {volume} {25}},\ \bibinfo {pages} {392001}
  (\bibinfo {year} {2013})}\BibitemShut {NoStop}%
\bibitem [{\citenamefont {Hohenadler}\ \emph {et~al.}(2011)\citenamefont
  {Hohenadler}, \citenamefont {Lang},\ and\ \citenamefont
  {Assaad}}]{Hohenadler11}%
  \BibitemOpen
  \bibfield  {author} {\bibinfo {author} {\bibfnamefont {M.}~\bibnamefont
  {Hohenadler}}, \bibinfo {author} {\bibfnamefont {T.~C.}\ \bibnamefont
  {Lang}},\ and\ \bibinfo {author} {\bibfnamefont {F.~F.}\ \bibnamefont
  {Assaad}},\ }\href@noop {} {\bibfield  {journal} {\bibinfo  {journal} {Phys.
  Rev. Lett.}\ }\textbf {\bibinfo {volume} {106}},\ \bibinfo {pages} {100403}
  (\bibinfo {year} {2011})}\BibitemShut {NoStop}%
\bibitem [{\citenamefont {Hohenadler}\ \emph {et~al.}(2012)\citenamefont
  {Hohenadler}, \citenamefont {Meng}, \citenamefont {Lang}, \citenamefont
  {Wessel}, \citenamefont {Muramatsu},\ and\ \citenamefont
  {Assaad}}]{Hohenadler12}%
  \BibitemOpen
  \bibfield  {author} {\bibinfo {author} {\bibfnamefont {M.}~\bibnamefont
  {Hohenadler}}, \bibinfo {author} {\bibfnamefont {Z.~Y.}\ \bibnamefont
  {Meng}}, \bibinfo {author} {\bibfnamefont {T.~C.}\ \bibnamefont {Lang}},
  \bibinfo {author} {\bibfnamefont {S.}~\bibnamefont {Wessel}}, \bibinfo
  {author} {\bibfnamefont {A.}~\bibnamefont {Muramatsu}},\ and\ \bibinfo
  {author} {\bibfnamefont {F.~F.}\ \bibnamefont {Assaad}},\ }\href@noop {}
  {\bibfield  {journal} {\bibinfo  {journal} {Phys. Rev. B}\ }\textbf {\bibinfo
  {volume} {85}},\ \bibinfo {pages} {115132} (\bibinfo {year}
  {2012})}\BibitemShut {NoStop}%
\bibitem [{\citenamefont {Rachel}\ and\ \citenamefont {LeHur}(2010)}]{Rachel}%
  \BibitemOpen
  \bibfield  {author} {\bibinfo {author} {\bibfnamefont {S.}~\bibnamefont
  {Rachel}}\ and\ \bibinfo {author} {\bibfnamefont {K.}~\bibnamefont {LeHur}},\
  }\href@noop {} {\bibfield  {journal} {\bibinfo  {journal} {Phys. Rev. B}\
  }\textbf {\bibinfo {volume} {82}},\ \bibinfo {pages} {075106} (\bibinfo
  {year} {2010})}\BibitemShut {NoStop}%
\bibitem [{\citenamefont {Soriano}\ and\ \citenamefont
  {Fernandez-Rossier}(2010)}]{Soriano}%
  \BibitemOpen
  \bibfield  {author} {\bibinfo {author} {\bibfnamefont {D.}~\bibnamefont
  {Soriano}}\ and\ \bibinfo {author} {\bibfnamefont {J.}~\bibnamefont
  {Fernandez-Rossier}},\ }\href@noop {} {\bibfield  {journal} {\bibinfo
  {journal} {Phys. Rev. B}\ }\textbf {\bibinfo {volume} {82}},\ \bibinfo
  {pages} {161302(R)} (\bibinfo {year} {2010})}\BibitemShut {NoStop}%
\bibitem [{\citenamefont {Lado}\ and\ \citenamefont
  {Fernandez-Rossier}(2014)}]{Lado}%
  \BibitemOpen
  \bibfield  {author} {\bibinfo {author} {\bibfnamefont {J.~L.}\ \bibnamefont
  {Lado}}\ and\ \bibinfo {author} {\bibfnamefont {J.}~\bibnamefont
  {Fernandez-Rossier}},\ }\href@noop {} {\bibfield  {journal} {\bibinfo
  {journal} {Phys. Rev. Lett.}\ }\textbf {\bibinfo {volume} {113}},\ \bibinfo
  {pages} {027203} (\bibinfo {year} {2014})}\BibitemShut {NoStop}%
\bibitem [{\citenamefont {Meng}\ \emph {et~al.}(2014)\citenamefont {Meng},
  \citenamefont {Hung},\ and\ \citenamefont {Lang}}]{meng}%
  \BibitemOpen
  \bibfield  {author} {\bibinfo {author} {\bibfnamefont {Z.~Y.}\ \bibnamefont
  {Meng}}, \bibinfo {author} {\bibfnamefont {H.~H.}\ \bibnamefont {Hung}},\
  and\ \bibinfo {author} {\bibfnamefont {T.~C.}\ \bibnamefont {Lang}},\
  }\href@noop {} {\bibfield  {journal} {\bibinfo  {journal} {Mod. Phys. Lett.
  B}\ }\textbf {\bibinfo {volume} {28}},\ \bibinfo {pages} {1430001} (\bibinfo
  {year} {2014})}\BibitemShut {NoStop}%
\bibitem [{\citenamefont {Yamaji}\ and\ \citenamefont {Imada}(2011)}]{Yamaji}%
  \BibitemOpen
  \bibfield  {author} {\bibinfo {author} {\bibfnamefont {Y.}~\bibnamefont
  {Yamaji}}\ and\ \bibinfo {author} {\bibfnamefont {M.}~\bibnamefont {Imada}},\
  }\href@noop {} {\bibfield  {journal} {\bibinfo  {journal} {Phys. Rev. B}\
  }\textbf {\bibinfo {volume} {83}},\ \bibinfo {pages} {205122} (\bibinfo
  {year} {2011})}\BibitemShut {NoStop}%
\bibitem [{\citenamefont {Zheng}\ \emph {et~al.}(2011)\citenamefont {Zheng},
  \citenamefont {Zhang},\ and\ \citenamefont {Wu}}]{DongZheng11}%
  \BibitemOpen
  \bibfield  {author} {\bibinfo {author} {\bibfnamefont {D.}~\bibnamefont
  {Zheng}}, \bibinfo {author} {\bibfnamefont {G.~M.}\ \bibnamefont {Zhang}},\
  and\ \bibinfo {author} {\bibfnamefont {C.}~\bibnamefont {Wu}},\ }\href@noop
  {} {\bibfield  {journal} {\bibinfo  {journal} {Phys. Rev. B}\ }\textbf
  {\bibinfo {volume} {84}},\ \bibinfo {pages} {205121} (\bibinfo {year}
  {2011})}\BibitemShut {NoStop}%
\end{thebibliography}

\providecommand{\noopsort}[1]{}\providecommand{\singleletter}[1]{#1}%

\end{document}